\documentclass[preprint,12pt]{elsarticle}
\usepackage{amssymb}

\def\bra{\,<\!} \def\ket{\!>\,} 
\begin{document}

\begin{frontmatter}

\title{Triaxial projected shell model study of chiral rotation in
odd-odd nuclei}


\author{ G.H.~Bhat$^{1}$, J.A.~Sheikh$^{1,2}$ and R.~Palit$^{3}$}
\address{$^1$Department of Physics, University of Kashmir, Srinagar,
190 006, India \\
$^2$Department of Physics and Astronomy, University of
Tennessee, Knoxville, TN 37996, USA\\
$^3$Department of Nuclear and Atomic Physics, Tata Institute of Fundamental Research, Colaba, Mumbai, 400 005,India}

\date{\today}

\begin{abstract}

Chiral rotation observed in $^{128}$Cs is studied using the newly developed microscopic
triaxial projected shell model (TPSM) approach. The observed energy levels and the
electromagnetic transition probabilities of the 
nearly degenerate chiral dipole bands in this isotope are well 
reproduced by the present model. This demonstrates the 
broad applicability of the TPSM approach, based on a schematic interaction and angular-momentum projection
technique, to explain a variety of low- and high-spin phenomena in triaxial rotating nuclei.

\end{abstract}



\end{frontmatter}

The classification of band structures from symmetry considerations has
played a central role in our understanding of nuclear structure
physics. Most of the rotational nuclei are axially symmetric with
conserved angular-momentum projection along the symmetry axis. This
symmetry has allowed to classify a multitude of rotational bands 
using Nilsson scheme and has been instrumental to unravel the intrinsic 
structures of deformed nuclei \cite{BM75,SN55}. 
Although, most of the deformed nuclei obey axial symmetry at
low-excitation energies and spin, there are also known regions of the
periodic table, referred to as transitional nuclei, that violate axial 
symmetry and are described using the triaxial deformed
mean-field. Further, some nuclei that are axial in the 
ground-state become triaxial 
at higher excitation energies and angular momenta.
There are several empirical observations indicating that
axial symmetry is broken in transitional regions. For instance, the
moment of inertia of the transitional nuclei rises sharply with
angular-momentum for the ground-state band that cannot be explained
using axial symmetry \cite{JS99}. 
Further, the transition quadrupole 
moment of the
ground-state band of transitional nuclei increases with spin and has
been explained using the triaxial mean-field model \cite{JY01}. 
We would
like to mention that this is still a debateable issue, these
features could also be explained using the $\gamma$-vibrational degree
of freedom rather than the stable $\gamma$-deformation used in the
above cited work. 

Recently, it has been shown that rotational motion of a triaxial deformed
mean-field can  also exhibit chiral symmetry with angular-momenta of valence
protons, valence neutrons and the triaxial core directed along the three
principle axes of the triaxial field. Chiral symmetry emerges from the
combination of triaxial geometry with an axis of rotation that lies out
of the three principle planes of the ellipsoid \cite{SF97}. It has been shown 
that 
nuclei in the mass $\sim$ 130 region exhibit chiral geometry with proton 
in the lower-half of the $h_{11/2}$ subshell with its angular-momentum
aligned along the short-axis,  neutron in the upper-half of the
subshell with angular-momentum along long-axis and the core
angular-momentum directed along the medium axis as the energy is minimum
for the triaxial shape along this axis \cite{TS04,TS03,KS01,AA01,SU03}. 
Till date, the best candidates for chiral doublet band structures are observed
in $^{128}$Cs and $^{135}$Nd isotopes \cite {EG06,MU07}. The measurement 
of the lifetime of the states
in the degenerate dipole bands plays an extremely crucial role in identifying
the chiral rotations in atomic nuclei \cite {PE06} and these
measurements have been performed for $^{128}$Cs.

Several theoretical approaches have been used to  probe the chiral
doublet band structures observed in odd-odd nuclei
 \cite{ SF01,PO04,VI00,SK02,PM03,SW06}. In most of the analysis, 
particle-rotor approach with valence proton and neutron coupled 
to the triaxial rotor has been employed. Although, this model has been
shown to provide a reasonable description of the most of the
characteristics of the chiral bands, it is known to depict
discrepancies with the observed data \cite{EG06}. For instance, the
measured B(E2) values
for the yrast band in $^{128}$Cs drop with spin and the
two-quasiparticle triaxial rotor (2QPTR)
calculations display opposite trend of increasing B(E2) with spin \cite{EG06}. 
The discrepancy is possibly related to the phenomenological nature of the
particle-rotor model approach with the triaxial core held fixed for all
spin values. In the present TPSM approach, the core or vacuum state has a 
microscopic structure as compared to 2QPTR where it is simply a triaxial rotor.

The purpose of the present work is to provide a detailed investigation
of the chiral rotation using
the microscopic triaxial projected shell model (TPSM) approach. This 
model has been recently used to investigate $\gamma$-vibrational and quasiparticle 
bands in even-even and odd-mass triaxial nuclei and it has been demonstrated to
provide an accurate description of the observed properties 
\cite{YK00,JG08,JG09,JG10,PJ02,YJ02}. In the present study,
we have generalized the TPSM approach for odd-odd mass and as a first application of this
development, the chiral rotation observed in $^{128}$Cs is investigated. In particular, the transition
probabilities, which contain the important information on the chiral geometry, are evaluated
in the present work.
The angular momentum projection 
technique used in the present model provides the
B(E2) and B(M1) transition probabilities between states with well defined angular momenta and a direct comparison with the measured values is feasible. 


The basis space of the TPSM approach for odd-odd nuclei, developed in the present work,
is composed of one-neutron and one-proton quasiparticle configurations:
\begin{equation}
\{ | \phi_\kappa \ket = {a_\nu}^\dagger {a_\pi}^\dagger | 0 \ket \}  .
\label{intrinsic}
\end{equation}
The above basis space is adequate to describe the ground-state
configuration of odd-odd nuclei. For the investigation of the excited
configurations, proton- and neutron- pairs need to be broken, which
shall form focus of our investigation in future.  
The triaxial quasi-particle (qp)-vacuum $ | 0 \ket $ in Eq.~(\ref{intrinsic}) is
determined by diagonalization of the deformed Nilsson 
Hamiltonian and a subsequent 
BCS calculations. This defines the  Nilsson+BCS triaxial qp-basis in the
present work. The number of basis configurations depend on the number of
levels near the respective Fermi levels of protons and neutrons. The 
configuration space is obviously large in this case as compared to the
nearby odd-mass nuclei, and usually several configurations contribute 
to the shell model wave function of a state with nearly equal
weightage. This makes the numerical results very sensitive to the 
shell filling and the theoretical predictions for doubly-odd nuclei become
more challenging. 

The states $ | \phi_\kappa \ket $ obtained from the deformed Nilsson 
calculations do not conserve rotational symmetry. To  
restore this symmetry, angular-momentum projection technique is applied. The effect of rotation
is totally described by the angular-momentum projection operator and the whole 
dependence of wave functions on spin is contained in the eigenvectors,
since the Nilsson quasi-particle basis is not spin-dependent.  
From each intrinsic state, $\kappa$, in (\ref{intrinsic}) a band is generated through 
projection technique. The interaction 
between different bands with a given spin is taken into account by diagonalising
the shell model Hamiltonian in the projected basis. 

The Hamiltonian used in the present work is
\begin{equation}
\hat{H} = \hat{H_0} - \frac{1}{2} \chi \sum_\mu  \hat{Q}_\mu^\dagger 
\hat{Q}_\mu - G_M \hat{P}^\dagger \hat{P} 
- G_Q \sum_\mu \hat{P}_\mu^\dagger \hat{P}_\mu,
\label{Hamilt}
\end{equation}
and the corresponding triaxial Nilsson Hamiltonian is given by
\begin{equation}
\hat H_N = \hat H_0 - {2 \over 3}\hbar\omega\left\{\epsilon\hat Q_0
+\epsilon'{{\hat Q_{+2}+\hat Q_{-2}}\over\sqrt{2}}\right\} ,
\label{nilsson}
\end{equation}
where $\hat{H_0}$ is the spherical single-particle shell model Hamiltonian,
which contains the spin-orbit force \cite{Ni69}. The second, third 
and fourth terms in Eq.~(\ref{Hamilt}) represent quadrupole-quadrupole, 
monopole-pairing, and quadrupole-pairing interactions, respectively. 
The axial and triaxial terms of the Nilsson potential in
Eq. \ref{nilsson} contain the parameters $\epsilon$ and $\epsilon'$, 
respectively, which are related
to the $\gamma$-deformation parameter by 
$\gamma$ = tan$^{-1}\frac{\epsilon'}{\epsilon}$. The strength of the 
quadrupole-quadrupole force $ \chi $ is 
determined in such a way that the employed quadrupole deformation $ \epsilon $
is same as obtained by the HFB procedure. 
The monopole-pairing force constants $G_M$ used in the calculations are
\begin{equation}
G_M ^\nu = \lbrack 20.12 - 13.13 \frac{N-Z}{A} \rbrack A^{-1}, ~~~~~ 
G_M ^\pi = 20.12 A^{-1} . 
\end{equation}
Finally, the quadrupole pairing strength $G_Q$ is assumed to be proportional to
the monopole strength,
$G_Q = 0.16 G_M$. All these interaction strengths are the same as those
used in our earlier
studies for the even-even nuclei in the same mass region \cite{JS99,JG08,JG09,JG10}. 
Thus, we have a consistent description for doubly-even and doubly-odd
nuclear systems. 

Once the projected basis is prepared, we diagonalize the Hamiltonian
in the shell model space spanned by $\hat {P}^{I}_{MK} | \phi_\kappa \ket $.
The projected TPSM  wave function is given by
\begin{equation}
|\sigma, I M \ket = \sum_{K,\kappa} f^\sigma_\kappa \hat{P}^I_{MK} | \phi_\kappa \ket .
\label{wave1} 
\end{equation}
Here, the index $\sigma$ labels the states with same angular momentum and $\kappa$
the basis states. In Eq. \ref{wave1}, $f^{\sigma}_{\kappa} $ are the weights of the basis state $\kappa$,   $\hat{P}^I_{MK}$ is three-dimensional angular momentum projection operator \cite{RS80}
 \begin{equation}
\hat P^I_{MK} = {2I+1 \over 8\pi^2} \int d\Omega\,
D^{I}_{MK}(\Omega)\, \hat R(\Omega).
\label{POp}
\end{equation}
Finally, the minimization of the projected energy with respect to the expansion coefficient,
$f^\sigma_{\kappa}$, leads to the Hill-Wheeler type equation
\begin{equation}
\sum_{\kappa '} (H_{\kappa \kappa'} - E_\sigma N_{\kappa \kappa'} ) f^{\sigma}_{
\kappa'} = 0 ,
\end{equation}
where the normalization is chosen such that
\begin{equation}
\sum_{\kappa \kappa'}f^\sigma_\kappa N_{\kappa \kappa'} f^{\sigma'}_{\kappa'}
= \delta_{\sigma \sigma'}.
\end{equation}                                                      

The angular-momentum-projected wave functions 
are in the laboratory frame of reference and can thus be directly used to compute 
the observables. 
The reduced electric quadrupole transition probability $B(E2)$ from an initial state 
$( \sigma_i , I_i) $ to a final state $(\sigma_f, I_f)$ is given by \cite {su94}
\begin{equation}
B(E2,I_i \rightarrow I_f) = {\frac {e^2} {2 I_i + 1}} 
| \bra \sigma_f , I_f || \hat Q_2 || \sigma_i , I_i\ket |^2 .
\label{BE22}
\end{equation}
In the calculations, we have used the effective charges of 1.6e for protons 
and 0.6e for neutrons. 
The reduced magnetic dipole transition probability
$B(M1)$ is computed by 
\begin{equation}
B(M1,I_i \rightarrow I_f) = {\frac {\mu_N^2} {2I_i + 1}} | \bra \sigma_f , I_f || \hat{\mathcal M}_1 ||
\sigma_i , I_i \ket | ^2 , 
\label{BM11}
\end{equation}
where the magnetic dipole operator is defined as  
\begin{equation}
\hat {\mathcal {M}}_{1}^\tau = g_l^\tau \hat j^\tau + (g_s^\tau - g_l^\tau) \hat s^\tau . 
\end{equation}
\begin{figure}[htb]
 \centerline{\includegraphics[trim=0cm 0cm 0cm
0cm,width=0.50\textwidth,clip]{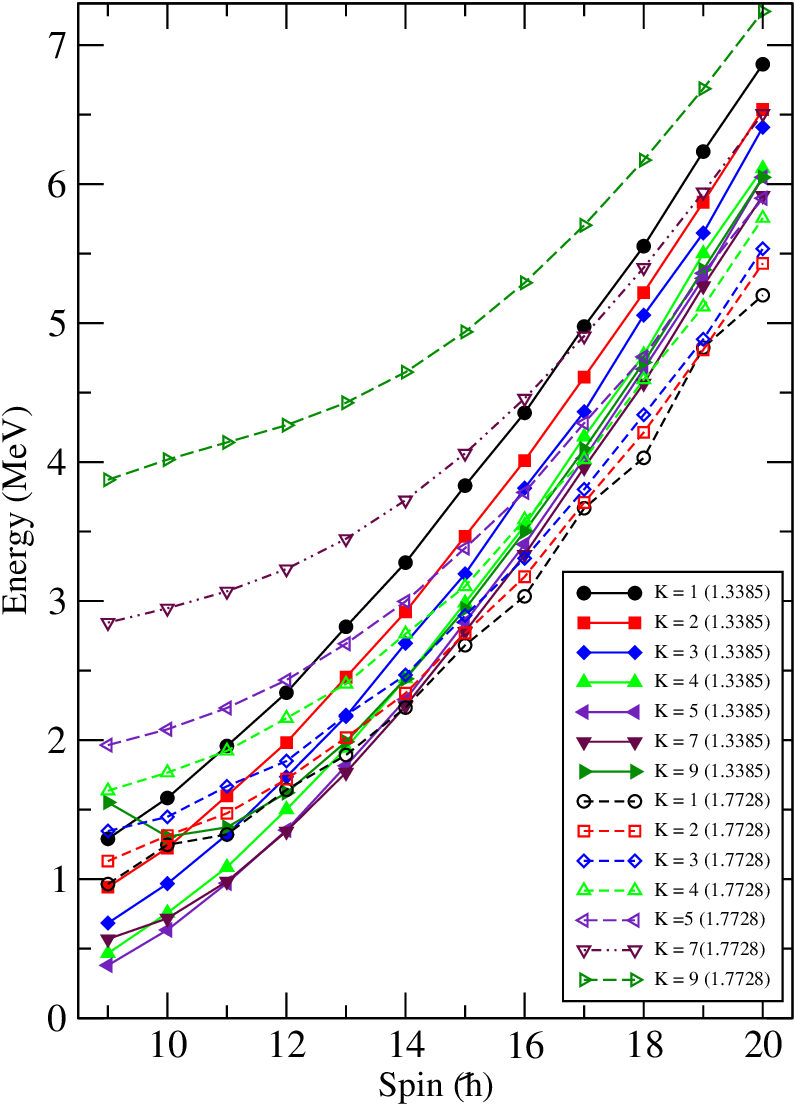}}
\caption{(Color online)  The angular-momentum projected bands obtained
 for different intrinsic K-configuration, given in legend box, for $^{128}$Cs nucleus. }
\label{fig1}
\end{figure}

\begin{figure}[htb]
 \centerline{\includegraphics[trim=0cm 0cm 0cm
0cm,width=0.50\textwidth,clip]{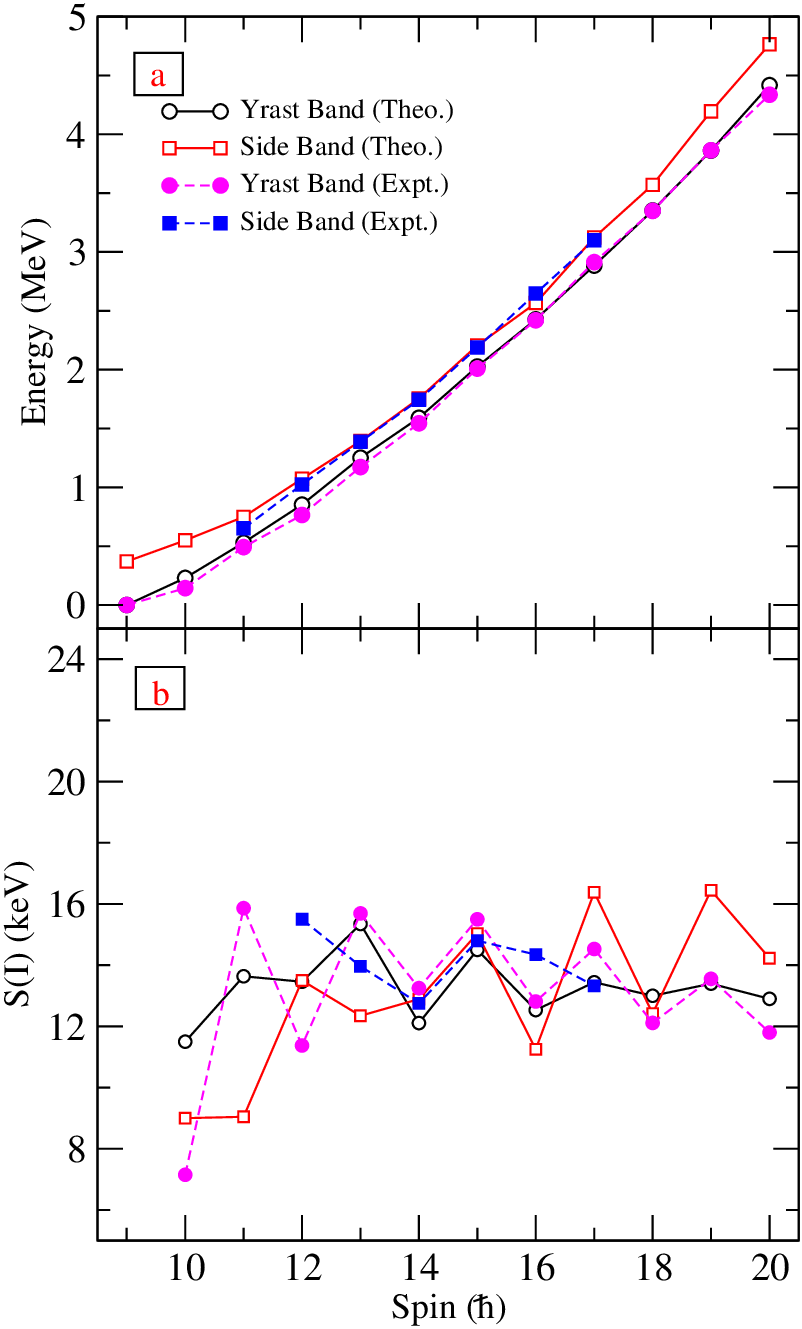}}
\caption{(Color online) Comparison of the TPSM energies after configuration 
mixing with the available experimental data for the yrast and side bands of the
studied $^{128}$Cs nucleus is shown in panel (a). Calculated staggering 
parameter $S(I)=\left[E(I)-E(I-1)\right]/2I $ is plotted along with the
 measured 
values in panel (b).}
\label{fig2}
\end{figure}
Here, $\tau$ is either $\nu$ or $\pi$, and $g_l$ and $g_s$ are the orbital and the spin gyromagnetic factors, 
respectively. 
In the calculations
we use for $g_l$ the free values and for $g_s$ the free values 
damped by a 0.85 factor, i.e.,
\begin{equation}
g_l^\pi = 1, ~~~ 
g_l^\nu = 0, ~~~   
g_s^\pi =  5.586 \times 0.85, ~~~ 
g_s^\nu = -3.826 \times 0.85.
\end{equation}
Since the configuration space is large enough to develop collectivity,
we do not use  core 
contribution. The reduced matrix element of an operator $\hat {\mathcal {O}}$ ($\hat {\mathcal {O}}$ is either
$\hat {Q}$ or $\hat {\mathcal {M}}$) is expressed by

\begin{eqnarray}
\bra \sigma_f , I_f || \hat {\mathcal {O}}_L || \sigma_i , I_i\ket    =  
\sum_{\kappa_i , \kappa_f} {f_{I_i \kappa_i}^{\sigma_i}} 
{f_{I_f \kappa_f}^{\sigma_f}}\nonumber & \\
 \sum_{M_i , M_f , M} (-)^{I_f - M_f}  
\left( \begin{array}{ccc}
 I_f & L & I_i \\
-M_f & M & M_i 
\end{array} \right) \nonumber 
   \bra \phi_{\kappa_f} | {\hat{P}^{I_f}}_{K_{\kappa_f} M_f} \hat {\mathcal {O}}_{LM}
\hat{P}^{I_i}_{K_{\kappa_i} M_i} | \phi_{\kappa_i} \ket & \\
\end{eqnarray}

In the present study, a detailed investigation of the chiral band structures observed in $^{128}$Cs
has been performed using the TPSM approach. We have chosen this nucleus since
it is the only system for which exhaustive lifetime measurements 
have been performed. As already discussed,
the electromagnetic transition probabilities are important to probe the chiral symmetry breaking of the observed dipole bands.   
In Fig.~1, the projected bands, obtained from the triaxially deformed intrinsic Nilsson
state by performing the three-dimensional angular-momentum projection,  are 
displayed. Nilsson intrinsic states have been obtained with deformation 
parameters, $\epsilon =0.220$ and $\epsilon'=0.14$.  The axial deformation value
has been adopted from particle-rotor model and other studies and the chosen 
value of non-axial deformation corresponds to $\gamma \sim 30^o$. The lowest 
bands
\begin{figure}[htb]
 \centerline{\includegraphics[trim=0cm 0cm 0cm
0cm,width=0.50\textwidth,clip]{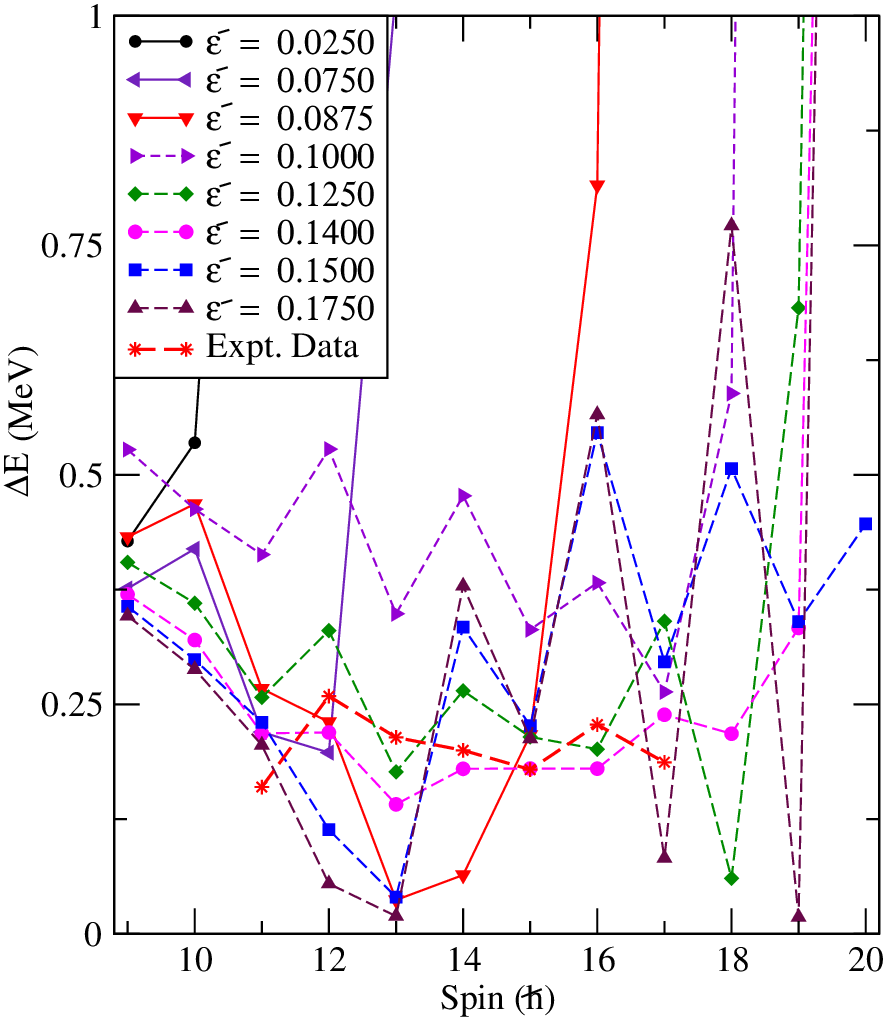}}
\caption{(Color online) Calculated energy difference ($\Delta E $)
 between the yrast and side bands as a function of angular-momentum for
 a range of the triaxial deformation $\epsilon'$ values. }
\label{fig2}
\end{figure}
\begin{figure}[htb]
 \centerline{\includegraphics[trim=0cm 0cm 0cm
0cm,width=0.50\textwidth,clip]{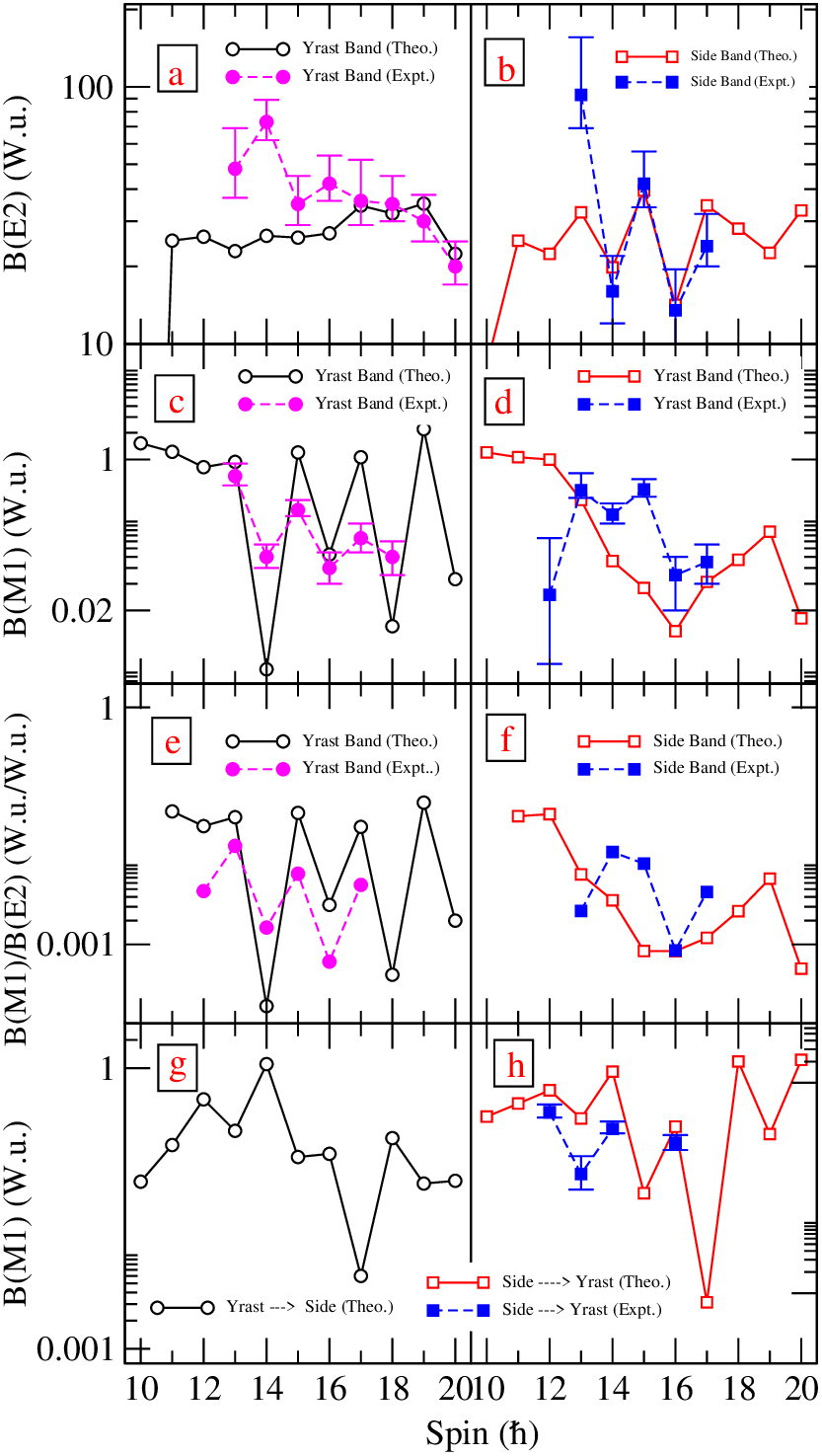}}
\caption{(Color online)  Comparison of the experimental B(E2), B(M1) 
transition strengths and their ratio for $^{128}$Cs \cite {EG06,HG10} with the 
calculated values for yrast and side bands are depicted in panels
(a) to (f). The calculated interband transitions are shown in panels (g)
 and (h), and panal (h) also depicts the measured values for a 
few transitions. }
\label{fig2}
\end{figure}
correspond to the projected states from the intrinsic configuration with
quasiparticle energy of 1.34 MeV and are obtained by specifying 
K-values in the three-dimensional angular-momentum projection operator, 
given in Eq.~(\ref{POp}). The projected bands obtained with K=4, 5 and 7 are degenerate
for I=9 and for higher spins up to I=13, K=5 and 7 are degenerate with
K=4 band becoming less favoured with increasing spin. For I=14, it is
noted that bands with K=1 and 2 belonging to the quasiparticle configuration
of energy 1.77 MeV cross the bands belonging to the quasiparticle energy
of 1.34 MeV. It is further observed that more bands belonging to the higher
quasiparticle energy cross at higher spins. This crossing of the 
quasiparticle states has not been explored in the earlier studies of the
chiral bands.

In the second stage of the calculations, projected bands 
obtained above are used to diagonalize the shell model Hamiltonian,
Eq.~(\ref{Hamilt}). It is important to mention that the projected
bands are not orthogonal, in general, and need to be orthogonalized 
before the diagonalizatoion of the Hamiltonian. The method that is adopted in the present approach, and in most 
of the generator coordinate methods, is to calculate the 
eigen-values of the norm matrix. The norm eigenvalues are non-negative 
quantities since the norm is a positive semi-definite matrix. 
However, it is quite possible that some of them may vanish. This 
happens under the circumstances that the 
multi-qp states become linearly dependent (non-orthogonal) when they
are projected. Such redundant states are removed simply 
by discarding the zero eigenvalue solutions of the norm eigen-value
equation. The details are given in the appendix of ref. \cite{KY95}.

Bands after diagonalization
have dominant compenents from several angular-momentum projected basis
configurations and the lowest two bands obtained are
displayed in the top panel of Fig.~2 along with the corresponding 
experimental chiral bands known for $^{128}$Cs. It is evident from Fig.~2(a) that
TPSM calculations reproduce the known experimental energies of the chiral
bands quite well. It is noted from the figure that the two bands are very close to each other up to I=14, and after this spin the two bands deviate with increasing spin. In Fig.~2(b), the staggering parameter, 
$S(I)=\left[E(I)-E(I-1)\right]/2I $ for the
two calculated chiral bands are plotted and compared with the corresponding
experimental numbers. It is noted that for low angular-momenta, the
calculated quantity deviates considerably from the experimental values,
but for higher angular-momenta a reasonable agreement is
obtained. Particle-rotor model results depict a similar discrepancy at
lower angular-momenta \cite{TS03,SF01}. 

To elucidate the importance of the triaxial deformation in the occurence
of degenerate dipole bands, in Fig. 3 the energy difference $(\Delta E)$
between the two dipole bands is drawn as a function of the the angular-momentum
for a range of the triaxial deformation values. It is evident from the
figure that for lower values of $\epsilon'$, close to the axial limit,
$\Delta E$ is quite large, especially for higher angular-momentum
states. However, for $\epsilon'$ larger than 0.125, it is noted that
$\Delta E$ becomes quite small for all angular-momentum states. 

We now discuss the electromagnetic transition probabilities of the
observed doublet bands in  $^{128}$Cs.
B(E2) (from I to I-2) and B(M1) (from I to I-1) transition probabilities 
have been calculated using the
expressions, Eqs. (\ref{BE22}) and (\ref{BM11}) and are shown in Fig.~4 for 
both the yrast and the side bands. 
B(E2) transition probabilities for the yrast band, displayed in Fig.~4(a),
depict a slight drop at around I=13 and is due to the crossing of the
quasiparticle state with energy equal to 1.77 MeV with the
ground-state quasiparticle configuration. The measured B(E2) for the 
yrast band, displayed in
Fig.~4(a), shows a more pronounced drop at a slightly higher spin value
of I=15. Further, the 
measured B(E2) after I=16 show a
smooth behaviour and is well reproduced by the TPSM 
calculations (see Table I). The calculated B(E2) for the
side band displays a little odd-even staggering, in particular, for
intermediate spin values. Except for I=13, the measured B(E2) for the
few transitions are in good agreement with the calculated values and 
is quite evident from Table I. 

The present TPSM approach provides a little improved description 
of the measured B(E2) values due to its microscopic 
treatment of the core (or vacuum) as compared to the phenomenological rotor 
core in 2QPTR. In TPSM, vacuum configuration (quasiparticle states) 
are generated by first solving triaxial Nilsson potential with the parameters 
$\epsilon$ and $\epsilon'$. In the present study, we have adopted $\epsilon=0.22$, 
which is, as a matter of fact, same as that employed in the particle-rotor 
model analysis. $\epsilon'= 0.14$ has been used as this corresponds to $\gamma \sim 
30^0$ and is important to lead to chiral configuration. In the second phase, 
standard BCS calculations are performed to obtain the quasiparticle
states with the Nilsson basis. Although, in TPSM angular-momentum projection is performed with pairing (monopole + quadrupole) plus quadrupole-quadrupole 
interaction for each angular-momentum, but due to limited quasiparticle space 
with only one-proton and one-neutron excitations considered, it is expected 
that deformation of the projected states will not be very different from 
the vacuum or mean-field configuration. In TPSM, the vacuum configuration or 
mean-field is held fixed for all the projected states and this is certainly a drawback of the present analysis. In a more accurate treatment, projection after 
variation, mean-field is re-calculated for each angular-momentum state. We 
consider that due to this problem in the present approach, we observe only a 
marginal drop in the B(E2) in comparison to the experimental data.


\begin{table}[t]
\caption{Experimental \cite {EG06} and calculated intra band B(E2) transition 
strengths (from $I\rightarrow I-2$) for yrast and side bands of $^{ 128}$Cs in $(W.u.)$ }
\begin{tabular}{ccccc} \hline
I$(\hbar)$ & B(E2)  &          & B(E2) & \\
           & Yrast Band &      &  Side Band & \\ \hline
           &  Experiment & Theory & Experiment & Theory \\ \hline
11  &  & 25.19   &  &  25.19      \\
12  &  & 26.10   &  &  22.37  \\
13  & $48^{+21}_{-11}$ & 22.94   & $93^{+63}_{-24}$ &  32.46       \\
14  & $73^{+16}_{-11}$ & 26.29   &$ 16^{+6}_{-4}$ & 19.84       \\
15  &$ 35^{+10}_{-6}$ & 25.90   & $42^{+14}_{-8}$  & 39.49    \\
16  & $42^{+12}_{-6}$ & 26.90   & $13.5^{+6.0}_{-5.9}$ & 14.18       \\
17  &  $36^{+16}_{-7}$ & 34.47 &   &   34.58      \\
18  & $35.0^{+10}_{-5}$ & 32.14 &  &   28.08     \\
19  & $ 30^{+8}_{-5}$ & 35.10 &   &   22.57       \\
20  & $20^{+5}_{-3}$ & 22.40 &  &   33.02   \\\hline
\end{tabular}
\end{table}

B(M1), shown in Fig.~4(c),  for the yrast band
are initially smooth, but after spin, I=13 show large odd-even
staggering. For the side band, B(M1) has slightly a different 
behavior with B(M1) transitions showing a drop for I=13 to 16. The
corresponding measured transitions are almost constant for I=13,14 and
15, and depict a drop for I=16.
The phase of the odd-even staggering and the 
magnitudes of the B(M1) transitions are quite similar in the 
two dipole bands and is one of the important criterion
for the chiralilty of the two bands. The calculated B(M1)/B(E2)
ratios, shown in Figs.~(4e) and (4f), are in good agreement with
the experimental values obtained from the intensities \cite{TS03}. 
The inter-band B(M1) transitions
between yrast to side and side to yrast, shown in Figs.~4(g) and 4(h),
have same phase of odd-even staggering and further are similar in
magnitude. What is important is that this phase is opposite to that of the intra-band B(M1)
transitions of Fig.~4(c) and 4(d) and forms another essential benchmark
for the existence of the chiral geometry \cite{TS04}. In Fig. 4(h), the
measured B(M1) values for a few available transitions are noted to be in
good agreement with the calculated values. 

In conclusion, the chiral dipole band structures observed in $^{128}$Cs have been investigated
using the triaxial projected shell model for odd-odd nuclei. It has
been demonstrated that observed energy levels and electromagnetic
transition probabilities of the dipole bands are quite well reproduced. 
Furthermore, it has been demonstrated that the 
calculated inter band B(M1) transitions from yrast to side and vice-versa, depict opposite staggering behaviour as compared to the the in-band transitions and are in conformity with the predicted properties of the chiral bands. 

The authors would like to acknowledge Dr. Ernest Grodner for providing
the measured data on the electromagnetic transitions. 


\end{document}